\documentclass[aps,epsf,subfigure,twocolumn]{revtex4}

\usepackage{graphicx}
\usepackage{dcolumn}
\usepackage{bm}
\usepackage{amsmath}
\usepackage{subfigure}
\usepackage{color,psfrag,epsfig}
\begin{document}

\newcommand{\up}{{\mid \uparrow \rangle}}
\newcommand{\down}{{\mid \downarrow \rangle}}

\title{Inversion symmetric two-level systems and the low temperature universality in disordered
solids}

\author{M. Schechter$^1$}
\author{P. C. E. Stamp$^{2,3}$}
\affiliation{$^1$Department of Physics, Ben-Gurion University of
the Negev, Beer-Sheva 84105, Israel}
\affiliation{$^2$Department of Physics and Astronomy, University of
British Columbia, Vancouver, British Columbia, Canada V6T 1Z1 }
\affiliation{$^3$Pacific Institute for Theoretical Physics, University
of British Columbia, Vancouver, British Columbia, Canada V6T 1Z1 }

\date{today}

\begin{abstract}

The low temperature universal properties in disordered and amorphous
solids are considered. We introduce a model that includes two types
of two level systems (TLSs), which, based on their local symmetry,
interact weakly or strongly with the phonon field. This
accounts well for the experimental results, and
addresses some long-standing questions: the nature of the TLSs; the
smallness and universality of the phonon attenuation, and the energy
scale of $3$K below which universality is observed. Our model
describes disordered lattices; we also discuss its application to
amorphous solids.

\end{abstract}

\maketitle

\section{Introduction}

Amorphous solids, and many disordered lattices, display
remarkably similar characteristics at low temperatures, which are very
different from those of ordered lattices \cite{zellerP}. Below a
rather universal temperature $T_{\rm U} \approx 3$K, their specific
heat $C_v(T)$ is nearly linear in temperature $T$, the thermal
conductivity $\kappa(T)$ is roughly $\propto T^2$ (both quantities
behave as $T^3$ in ordered lattices), and their internal friction
$Q$ is independent of $T$ and of the phonon wavelength $\lambda$.
Moreover, the thermal conductivity and internal friction vary little
between materials whose microscopic structure ranges from impurities
in ordered crystals to completely disordered amorphous glasses. This
implies a rather universal ratio between the phonon mean free path
$l$ and the phonon wavelength $\lambda$, so that $l/\lambda \approx
150$ \cite{zellerP,urbana,HR86,pohl02}, and suggests that a
fundamental mechanism may dictate the low-$T$ behavior of disordered
solids. What this mechanism might be is the 'universality problem',
which has emerged as one of the outstanding unsolved mysteries in
condensed matter physics \cite{pohl02,AJL88}.

Most theoretical analyses begin with the influential "Standard
Tunneling" (ST) model \cite{ABV72,Phillips72}. In this model the
low-energy modes are described by a single set of non-interacting
localized two-level systems (TLSs), which interact weakly with phonons.
These TLSs are characterized by two parameters, viz.; $\epsilon$ (the
energy bias between the wells), and $\Delta_o$ (the tunneling
amplitude between the wells). For an ensemble of such TLSs, the ST
model assumes a broad joint probability distribution for $\epsilon$ and $\Delta_o$
\cite{ABV72,Phillips72}, given by $P(\epsilon,\Delta_o) =
P_o/\Delta_o$, where $P_o$ is a constant. The central dimensionless
parameter of the ST model is the "tunneling strength" $C_o = P_o
\gamma^2/\rho c^2$, where $\gamma$ is the defect-phonon coupling,
$\rho$ the mass density, and $c$ the phonon velocity. The ST model
gives results agreeing well with the temperature dependence of
$C_V(T)$, $\kappa(T)$, $Q(T)$, and $l(T)$
\cite{ABV72,Phillips72,Jackle72}, if $C_o$ is assumed to be an
adjustable parameter. Universality would then be found if one
assumed that $C_o \approx 10^{-3}$ for the wide range of amorphous
and disordered systems.

The ST model leaves some central questions
open, viz.: (i) What is the nature of the TLSs? (ii) Why
is $C_o$ so small and universal? (iii) What dictates the energy
scale of $3$K below which universality is observed? (iv) what about
inter-defect interactions, which are not small? Thus, the ST model
has been widely questioned, and the problem of universality has been the
subject of thorough theoretical investigation
\cite{klein78,AJL88,BNOK98,parshin94,wolynes01,sethna85,solfK94,Kuhn03,parshin07}.

In disordered lattices universal properties have also been observed in the ferroelectric
phase \cite{MKS+85,YKMP86}, and recent experiments on mixed crystals suggest that
strong random strain fields acting on tunneling defects may be
responsible for universality \cite{watson95,pohlT99,pohlT02}.
Furthermore, experiments on ion-implanted crystalline Si show that
universality is not related to
amorphicity\cite{LVP+98}. This has led to the argument \cite{pohl99}
that experimental and theoretical studies on universality should
focus on disordered crystals, rather than on amorphous systems.

In this paper, we present an approach to the problem, relating
the universality to the symmetry of TLS states under local
inversion. We will begin by considering a strongly-disordered
crystal (not yet fully amorphous) possessing off-center or
rotational impurities, each having $2N$ local states between
which it can tunnel. This system has 2 kinds of
low-energy excitations (symmetric and asymmetric under inversion),
coupling very differently to phonons. At low energies we find that
only the symmetric excitations are active, and that a quantitative
understanding of universality can then be obtained. We also argue
that this model has a natural generalization to fully amorphous
systems, also leading to universality.

\begin{figure*}
\subfigure[]{
  \includegraphics[width=0.75\columnwidth]{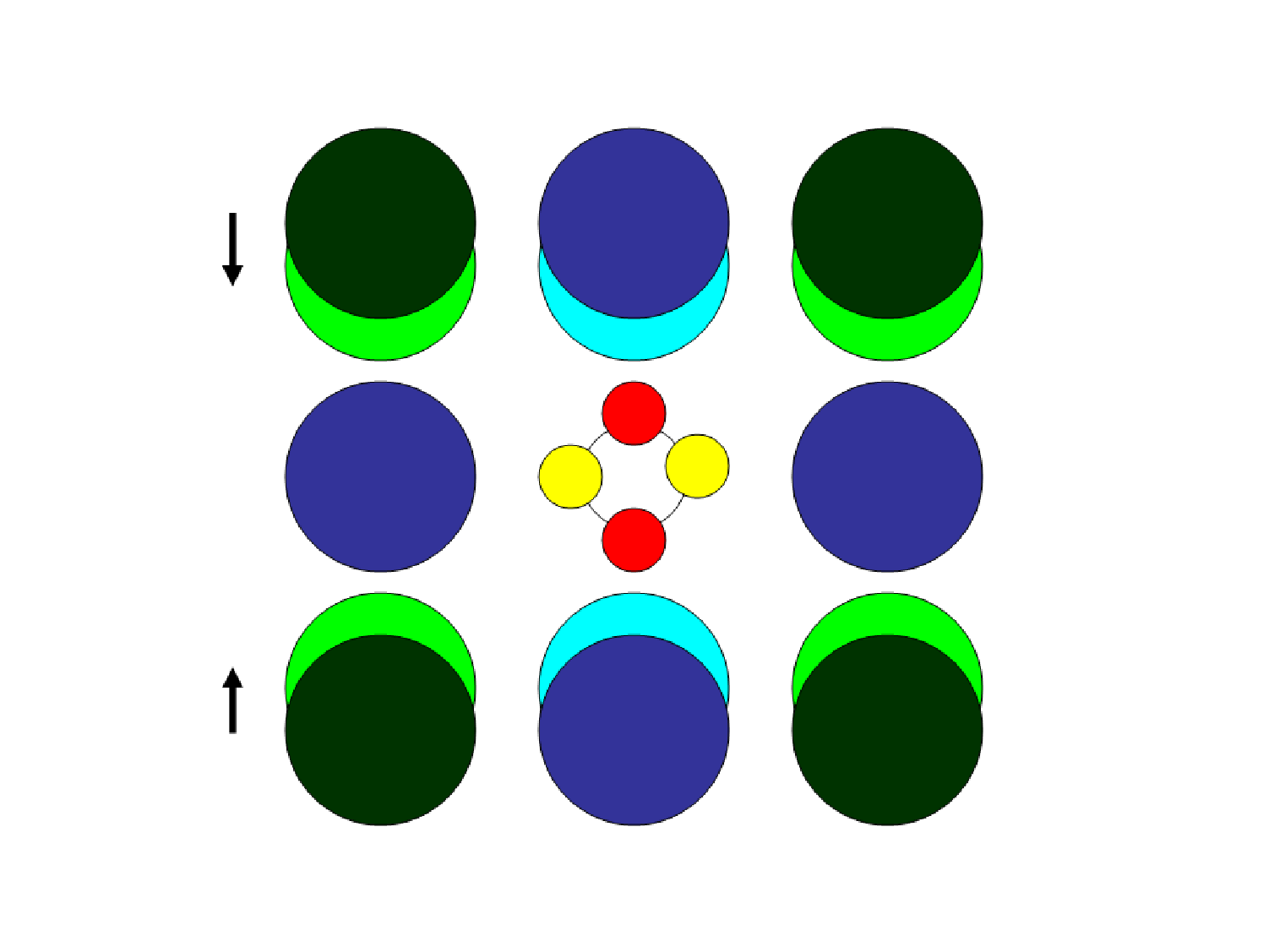}
   \label{fig:Fig1subfig1}
   }
 \subfigure[]{
  \includegraphics[width=0.75\columnwidth]{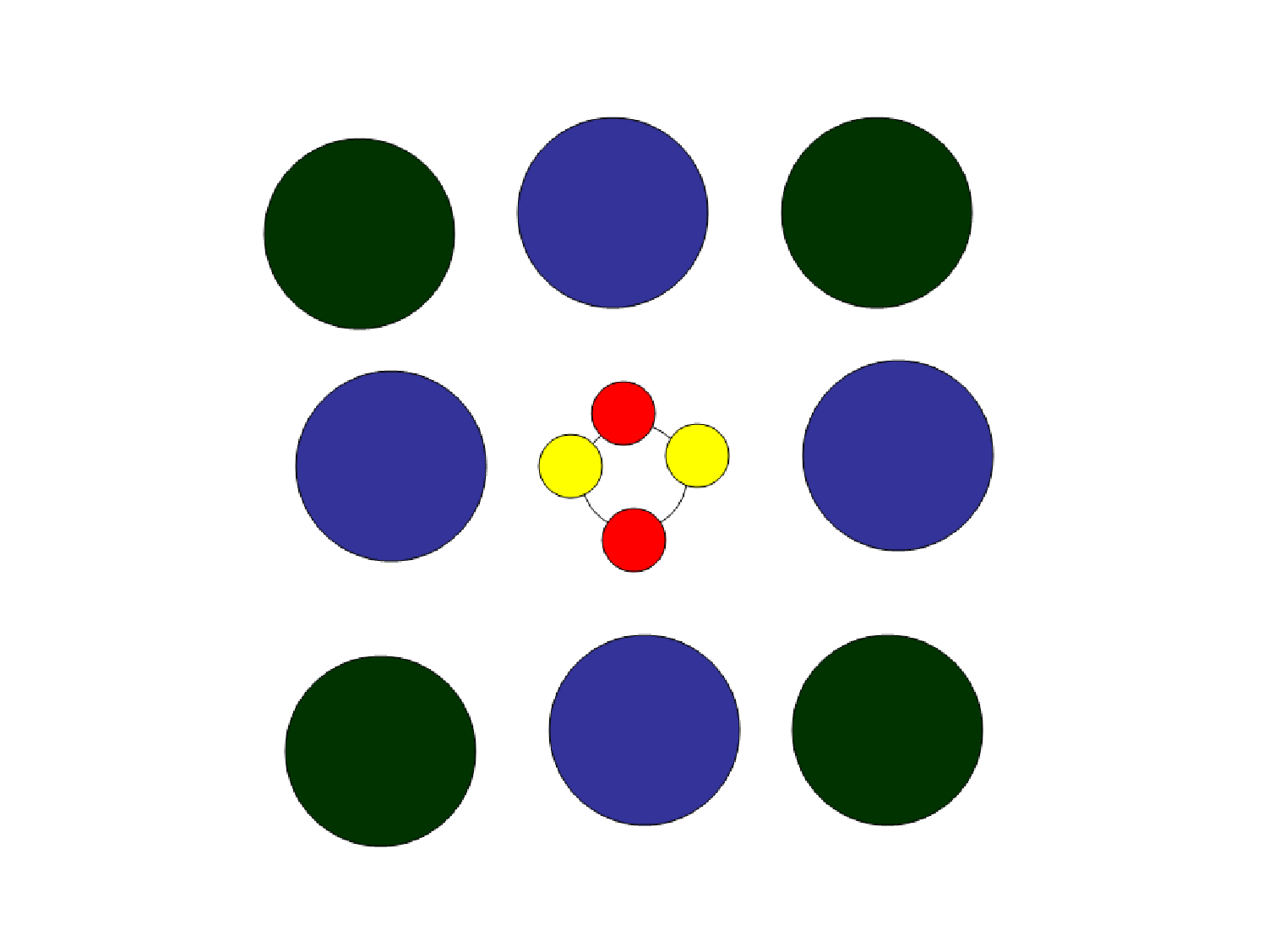}
   \label{fig:Fig1subfig2}
   }
 \subfigure[]{
  \includegraphics[width=0.52\columnwidth]{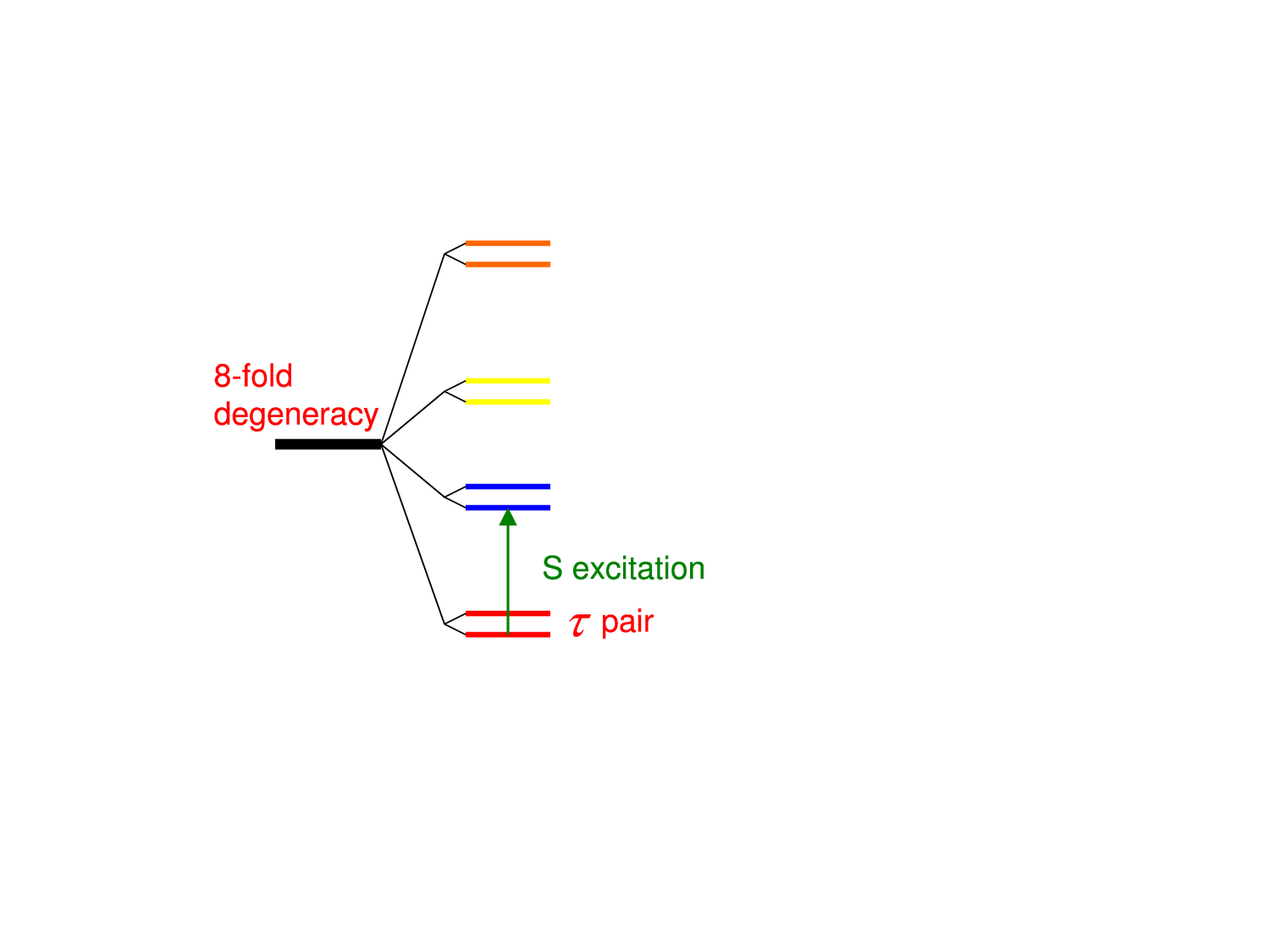}
   \label{fig:Fig1subfig3}
   }
   \hspace{2cm}
 \subfigure[]{
  \includegraphics[width=0.75\columnwidth]{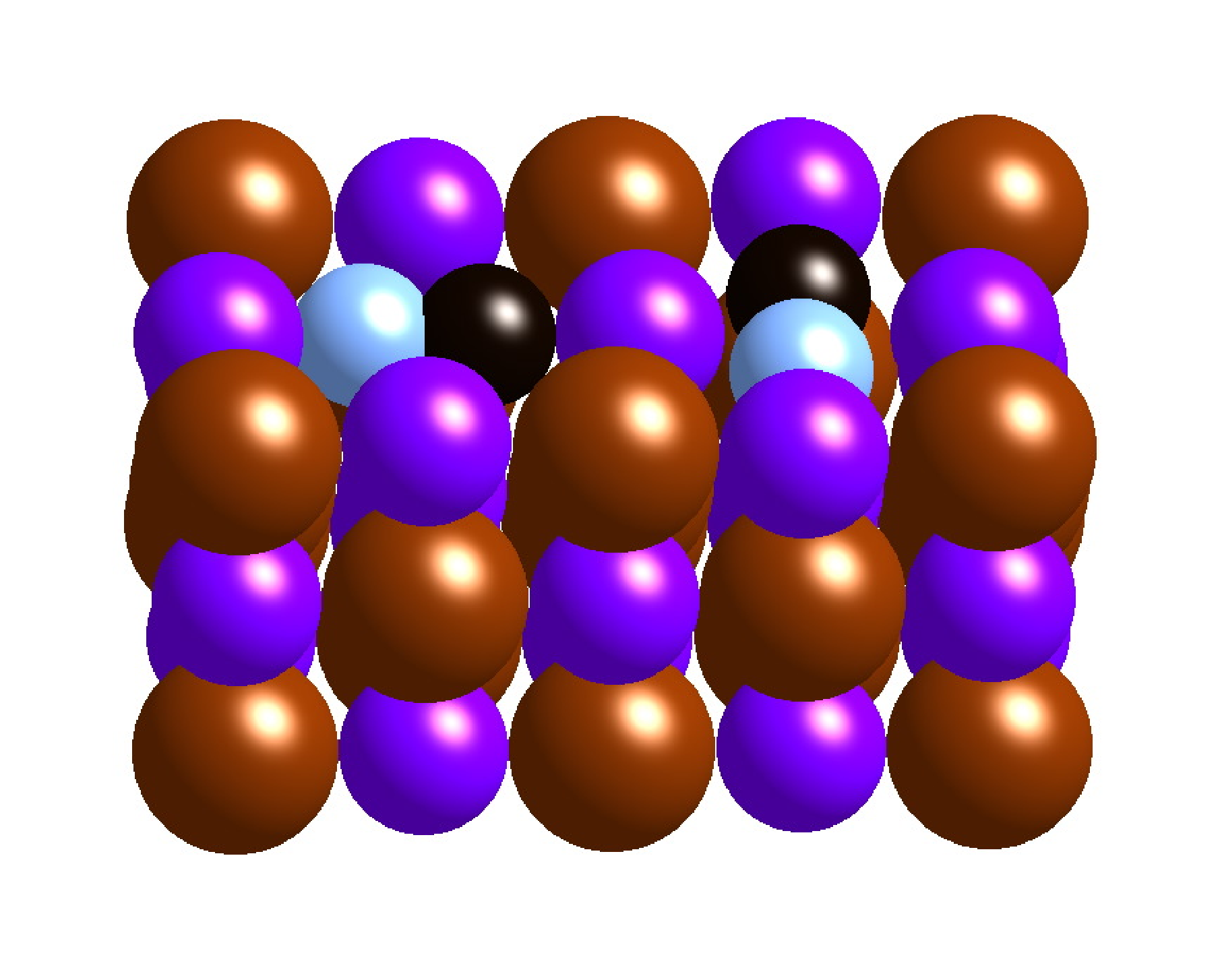}
   \label{Fig:Fig1subfig4:}
   }
\caption{$S$- and $\tau$-spins, and their interaction with elastic
strain.
(a) A 2-dimensional system in which
the impurity can occupy one of four positions
shown by the red and yellow circles; states of the same color are
related by inversion symmetry and form a "$\tau$-pair".
Different colors correspond to different S-subspaces, between
which transitions are asymmetric with respect to inversion.
Distortion by a passing phonon breaks the degeneracy between states of different colors (ie.,
couples to the $S$-excitations) but not between states of the same
color (i.e., does not couple to the $\tau$-excitations).
(b) The $\tau$-pair degeneracy is broken in a system with finite
strain created by strong lattice disorder or amorphousness. This results
in a finite but small $\tau$-TLS lattice interaction $\gamma_{\tau}$.
(c) We show $2N=8$ states of an impurity, initially assumed degenerate, which are then split in a disordered crystal. The splitting between states in the same $\tau$ pair is small. (d)
A specific example: the 3-dimensional alkali halide $KBr$ with
two $CN$ impurities, which, in strong disorder, align along one of the three crystal axes
(distorting the nearby lattice). Different $\tau$ states are connected by
$180^o$ $CN$ flips, whereas different $S$ states are connected by $90^o$ rotations.}
    \label{fig:tauS}
\end{figure*}

The paper is organized as follows: in Sec.\ref{sec:model} the model is introduced. In Sec.\ref{Experimental} we show how our model leads to universality, and explains other long standing experimental observations. We conclude in Sec.\ref{sec:conclusions}. The technical derivation of our model from microscopic considerations is detailed in App.~\ref{Sec:Mapping}. In App.~\ref{SEC:gapdos} we describe in detail the derivation of the density of states of the symmetric and asymmetric TLSs, in the absence and presence of random fields.

\section{The model}
\label{sec:model}

Many disordered crystals showing universal low-$T$
properties have defects that can be modeled as TLSs that are either symmetric
or asymmetric under local inversion.  Some examples are
(i) CN flips and rotations in KBr:CN
(Fig. 1(d)); (ii) F tunneling between interstitial states and between a
lattice position and a vacant interstitial position in ${\rm
CaF_2{:}LaF_3}$ \cite{CP89}; and (iii) double N2 rotations and
single N2 rotations in ${\rm ArN_2}$\cite{GS11,GAS12}.

\subsection{Effective Hamiltonian}

Consider first
a {\it single} off-center or rotational impurity at lattice site $j$. In
an otherwise pure lattice this impurity can occupy an even number $2N$ of states
at different points in the lattice cell
- pairs or 'doublets' of states related by inversion symmetry are classically degenerate (this degeneracy being weakly lifted by tunneling between sites \cite{tunnelingnote}).
We model the impurity or defect by treating each doublet as a 'pseudo-spin' $\hat{\bm{\tau}}_{jn}$, where $n =
1,2,..N$ labels the different doublets, and the Ising variable $\hat{\tau}_{jn}^z = |jn \uparrow
\rangle \langle jn \uparrow| - |jn \downarrow \rangle \langle jn \downarrow|$ describes the
'polarization' between the pseudospin states. We likewise define an Ising variable
$\hat{S}_{jnn'}^z = |jn \rangle \langle jn| - |jn' \rangle \langle jn'|$, describing the
polarization between different doublets $|jn \rangle$ and $|jn'\rangle$ at site $j$ (compare Fig. 1(a)). Note that any given doublet couples very weakly to phonons (only to the gradient of the strain field \cite{MS08}), because of the inversion symmetry.

Now consider a system with some concentration $x$ of randomly distributed impurities, so that the typical distance between impurities is $R_o = a_ox^{-1/3}$ where $a_o$ is the interatomic distance. The lattice strain generated by the impurities break the inversion symmetry, and scramble the states at each site (Fig. 1(b), 1(c)). The effective  low energy Hamiltonian of the disordered defect system can then be written as \cite{halp77,MS08}

\begin{eqnarray}
\hspace{-0.5cm}
H_{S \tau} &=& \sum_j [h_j^S S_j^z + h_j^{\tau} \tau_j^z] \nonumber \\
&+& \sum_{ij} [J_{ij}^{SS} S_i^z S_j^z  +
J_{ij}^{S\tau} S_i^z \tau_j^z + J_{ij}^{\tau \tau} \tau_i^z
\tau_j^z]
 \label{H-St}
\end{eqnarray}
where the interaction strengths $h^{a}, J^{ab}$ (with $a,b = S,\tau$) are random variables, and only the two low energy pairs are considered, (see details in App.~\ref{Sec:Mapping}).

The size of the random couplings follows a simple rule, according to which

\begin{equation}
J_{ij}^{ab} = \frac{c_{ij}^{ab} {\gamma}_a {\gamma}_b}{\rho c^2 (R_{ij}^3+ \tilde{a}^3)}
 \label{hJ-mag}
\end{equation}
where $R_{ij}$ is the inter-defect distance, $\tilde{a}$ is a short distance cutoff for the
interaction, $\gamma_S$ and $\gamma_{\tau}$ are the phonon couplings to the $S$ and $\tau$ defects {\it in the presence of the disorder}, and the randomness in the magnitude and angular dependence \cite{halp77,MS08} of these couplings are absorbed into the random variables
$c_{ij}^{ab} \sim O(1)$. Thus the three interactions have the same radial dependence; however their energy scales are quite different. At a distance $R_{ij} = R_0$, one finds typical interaction strengths $J_o   \equiv J^{SS} \propto \gamma_S^2 \sim 500$K, $J^{S\tau}
\propto \gamma_S \gamma_{\tau} \sim g J_o \sim 10$K and $J^{\tau \tau} \propto \gamma_{\tau}^2
\sim g^2J_o \sim 0.2$K. The random field strengths are also governed by $J_o$ and $g$; their typical strengths \cite{MS08} are $h^S \lesssim J_o$ and $h^{\tau} \lesssim g J_o$.

A key feature of these results is the role of the coupling $\gamma_{\tau}$ to the phonons. Unlike the situation in the absence of disorder, where the inversion symmetric $\tau$-TLSs couple only to the strain gradient, strong disorder destroys inversion symmetry, and results in a finite coupling $\gamma_{\tau}$ to the strain (see Eq.~(\ref{V-cont-disorder}) and the following derivation of Eq.~(\ref{H-St}) in App.~\ref{SEC:TLSPhonon}).
Since the coupling $\gamma_{\tau}$ is a result of the rather small deviations from inversion symmetry, it is much smaller than the coupling of the asymmetric $S$-TLSs to the strain.
The ratio $g = \gamma_{\tau}/\gamma_S$ can be determined microscopically (see App.~\ref{SEC:TLSPhonon}) and also estimated by simple dimensional arguments: in strongly disordered systems the typical displacement of a lattice site due to the random lattice strain will be $\delta a \approx a_o (E_{\Phi}/E_C)$, where $E_{\Phi}$ is the Debye energy characteristic of acoustic phonons, and $E_C$ is the typical Coulombic correlation energy in the solid. Thus we expect $g \approx \delta a/a_o \approx E_{\Phi}/E_C \sim 0.02 $ (see also Refs.\cite{GS11,CBS13}).

A mean-field (MF) treatment of (\ref{H-St}) yields a picture of independent $S$ and $\tau$ variables, acted upon by the random couplings in the Hamiltonian (\ref{H-St}). This spreads out the states, resulting in Gaussian mean field
densities of states
$n^o_S(E)$, $n^o_{\tau}(E)$, of width $J_o, gJ_o$ and peak height $\propto 1/J_o,
1/gJ_o$, for the $S$ and $\tau$ excitations respectively (see Fig. 2(a)). In this MF picture $n^o_{\tau}(E) \gg n^o_{S}(E)$ for $E
\ll g J_o$, by a factor $1/g$; nevertheless, the $S$-spins dominate the phonon scattering even at
low energies, because their scattering rate is $\propto \gamma_S^2$, and thus far higher (by a factor
$1/g^2$) than that for $\tau$ spins.

Thus the model of $S$ and $\tau$ pseudospins gives a nice simple picture in MF theory.  However,
MF theory is misleading - it neglects correlations between the $S$ and $\tau$ variables, which
radically change the low-energy physics.

\subsection{$S$-$\tau$ Correlations}

The correct way to handle these correlations is using an
Efros-Shklovskii treatment \cite{efros75}, adapted here for the case of two types of interacting
TLSs and the Hamiltonian (\ref{H-St}). The details are technical (see App.~\ref{SEC:gapdos}),
but the physics of the result is straightforward (Fig. 2(a)). Level repulsion between the 2 sets of variables
has a radical effect on $n_S(E)$, but little effect on $n_{\tau}(E)$, simply because the MF level
density $n_{\tau}(E)$ is much higher. Thus, $n_S(E)$ shows a slow fall-off for $g J_o < E < J_o$,
but a precipitous drop for $E < g J_o \sim 10$K, so that when $E \ll gJ_o$, the $S$-states have
essentially disappeared. However $n_{\tau}(E)$ only shows a weak dip below the much lower energy
$J_o^{\tau \tau} \sim g^2 J_o \sim 0.2$K, caused by $J_o^{\tau \tau}$. The phonon spectrum is
only weakly altered. These results, shown in more detail in Fig. 2(b)-2(d), are also found in Monte Carlo calculations for the Hamiltonian in Eq.(\ref{H-St}) \cite{CGBS13} and in a hybrid molecular statics and Monte Carlo calculation for the KBr:CN system using only bare interatomic potentials \cite{CBS13b}.

This abrupt switch in the DOS, from $S$-states to $\tau$-states, has a crucial consequence - there is a crossover in the
system properties at a temperature $T_U$ (defined by the condition $\gamma_{\tau}^2 n_{\tau}(T_U)
= \gamma_{S}^2 n_S(T_U)$). The crossover temperature $T_U \approx 0.2 g J_o$, and is only
weakly dependent on $\tilde{a}$ for typical values $\tilde{a}/a_o \sim 1-6$.
Above $T_U$, the $S$-pseudospins predominate, along with a plentiful supply of phonons, to which
they couple strongly, thereby dominating the phonon attenuation. In contrast, below $T_U$ the $S$
spins are frozen and exert on the $\tau$ TLSs
a random field much larger than the $\tau \tau$ interaction $J^{\tau \tau}$. The
$\tau$-pseudospins then behave like a set of non-interacting TLS states in strong disorder;
However, they now couple more strongly to phonons than the $S$ pseudospins, and dominate the
phonon attenuation. In the fairly narrow crossover regime around $T_U$, the $S-\tau$
interactions determine the detailed shape of the crossover.

\begin{figure*}
\begin{center}
\subfigure[]{\includegraphics[width=0.8\columnwidth]{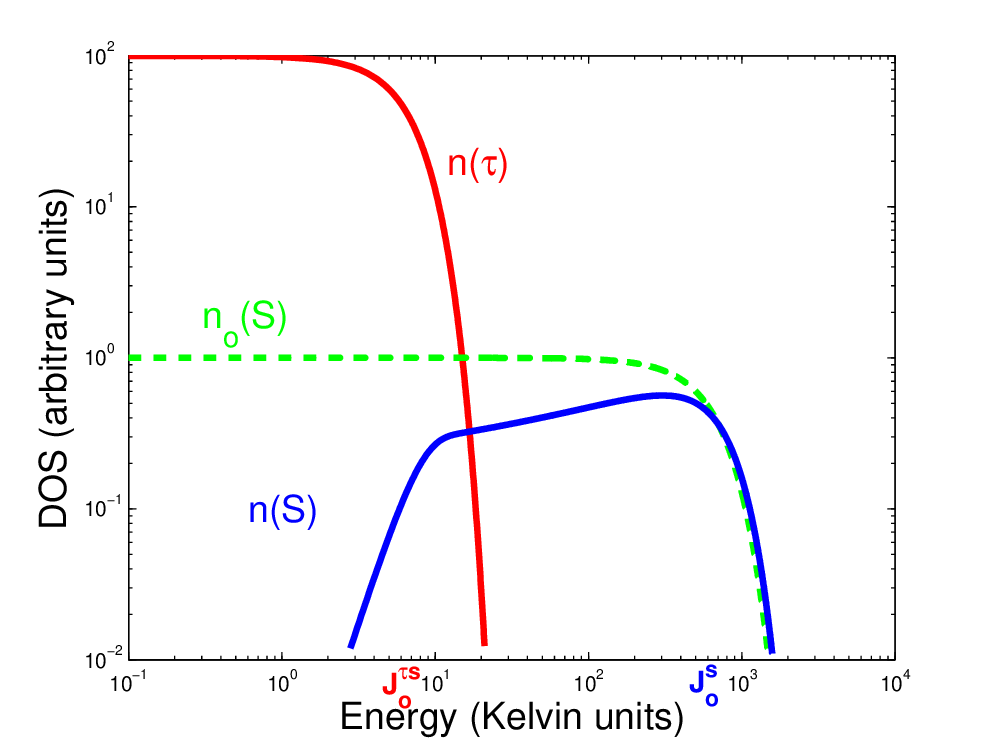}}
\subfigure[]{\includegraphics[width=0.8\columnwidth]{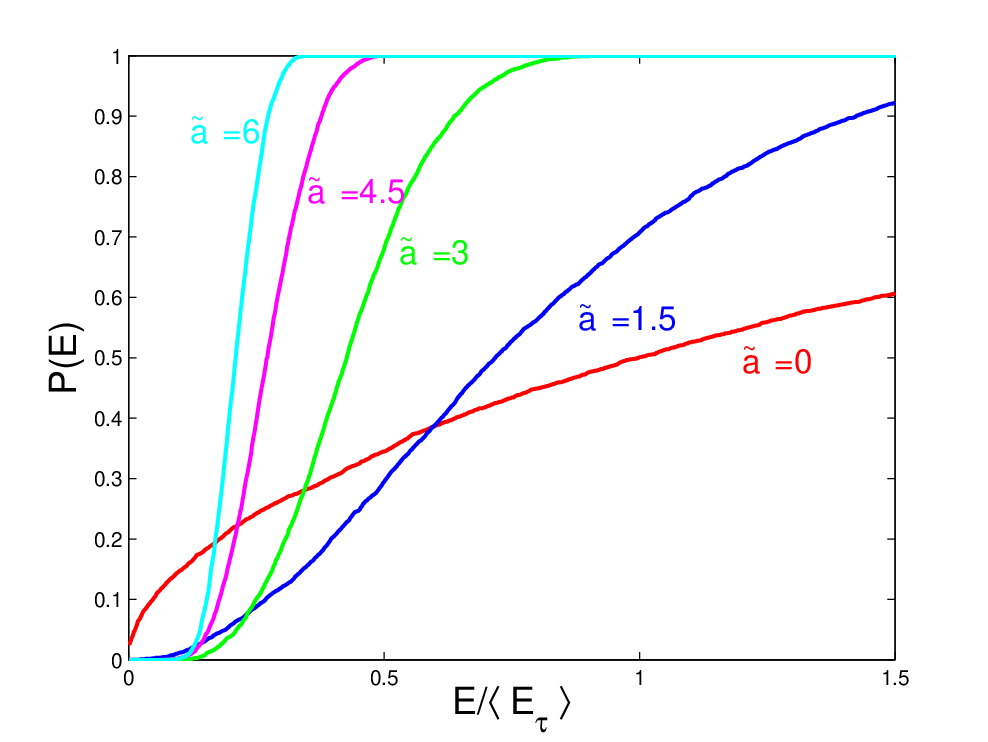}}
\subfigure[]{\includegraphics[width=0.8\columnwidth]{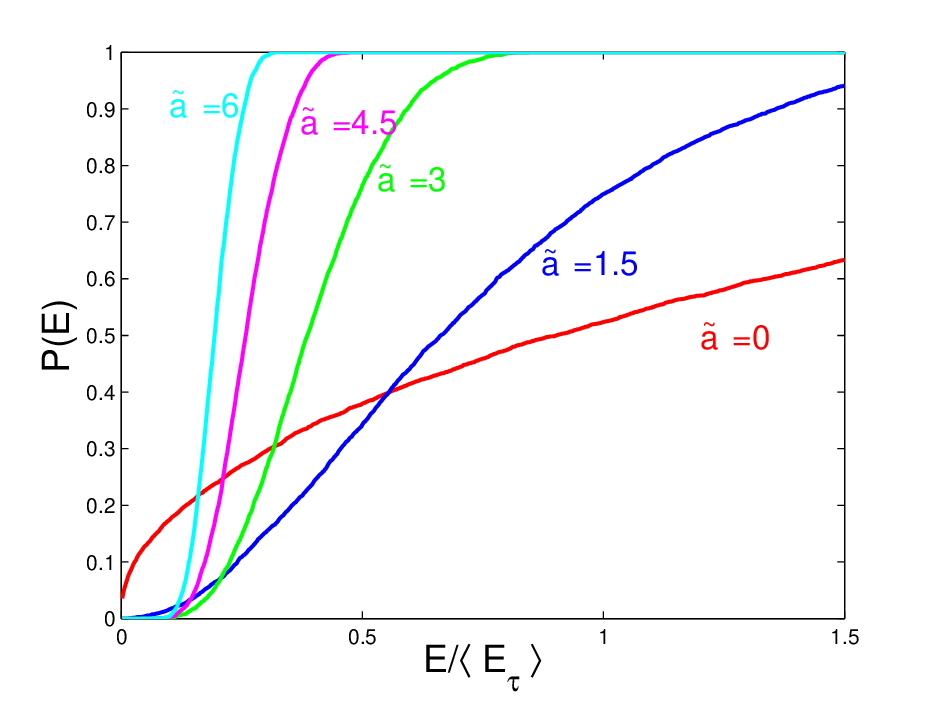}}
\hspace{0.2cm}
\subfigure[]{\includegraphics[width=0.8\columnwidth]{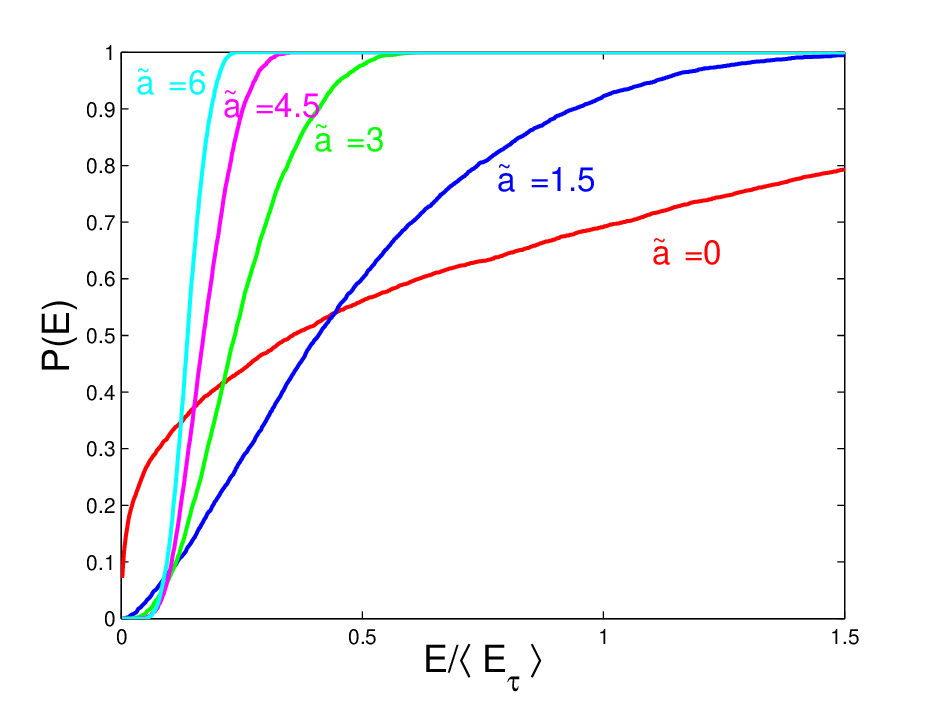}}
\end{center}
\caption{Densities of states (DOS) of $\tau$ and $S$ TLSs.
(a) We show $n_{\tau}(E)$ in red (it is hardly changed by $S-\tau$ interactions) and
$n_S(E)$ in blue (the non-correlated DOS $n^o_S(E)$ is shown in green for comparison).
In (b), (c), (d) we show the reduction factor $P(E) \equiv n_S(E)/n_S(\bar{E})$, where $\bar{E}$ is an energy a few times
larger than $g J_o$; essentially $n_S(\bar{E})$ is the $S$-spin
DOS obtained if one includes $S-S$ correlations but
ignores $S-\tau$ correlations, so $P(E)$ measures the effect of
$S-\tau$ correlations. Numerical results are shown for
$\tilde{a}=0;1.5;3;4.5;6$, for an impurity concentration
x$=0.2$ and sample size $18^3$
cells (i.e., $\sim 4650$ TLSs), in the 3 cases
(b) $h^{\tau}=0$;
(c) $h^{\tau}/\langle E_{\tau} \rangle \approx 0.3$; and
(d) $h^{\tau}/\langle E_{\tau} \rangle \approx 1. \,\,\,$
$T_U$ is defined by the energy at
which $n_{\tau}(E) \gamma_{\tau}^2 = n_S(E) \gamma_{S}^2$, i.e., by $P(E) \approx 5g$.
Similar results for $x = 0.5$ are shown in Fig. \ref{dos05}.
For both concentrations we find $T_U \approx 0.2 \langle E_{\tau} \rangle $
for $1 < a_o < 6$, i.e., $T_U \approx 0.2 g J_o$.}
    \label{Staugap}
\end{figure*}

\section{Relation to experiments}
\label{Experimental}

Our model above, microscopically derived for the disordered lattices, results in the dominant TLSs at low temperatures having a nearly homogenous DOS at low energies, and a TLS-TLS interaction which is much smaller than their random energies. Furthermore, the fact that these tunneling systems consist of two levels, assumed in the ST model, and later observed experimentally\cite{HR86}, is intrinsic in our model, a result of inversion symmetry. In that, our model is similar to the ST model. However, the central advance in our model with respect to the ST model is that within our model there exists a generic relation between the coupling of the TLSs to the strain and their low energy DOS. This, along with the smallness of the coupling of the $\tau$-TLSs to the strain, and the fact that they are not gapped at low energies, allows us to explain some long standing experimental results, above all the universality and smallness of phonon attenuation, and the energy scale of $3$K below which universality is observed.

\subsection{Universality}
\label{sec:universality}

The universality is an immediate consequence of the above results, and is
found by simply adapting the original ST model analysis to the system of $S$ and $\tau$ spins. Let's first consider the strong disorder regime, where $x \sim O(1)$.
The DOS of the $\tau$-TLSs is then given by $n_{\tau} \approx (1/g J_o R_0^3) = \rho c^2/\gamma_{\rm
\tau} \gamma_{\rm S}$. At the same time, by integrating $P(\epsilon,\Delta_o)$ over $\Delta_o$ one obtains
$P_o = \kappa n_{\tau}$, with $\kappa^{-1} \equiv ln(\Delta_o^{\rm u}/\Delta_o^{\rm l})$, where
$\Delta_o^{\rm l},\Delta_o^{\rm u}$ are the lower and upper
cutoffs for the 'tunneling amplitudes' of the $\tau$ TLSs. We thus find
\begin{equation}
C_o = \frac {\kappa n_{\tau} \gamma_{\rm
w}^2}{\rho c^2} \approx \frac {\kappa \gamma_{\tau}}{\gamma_{S}}
\approx \kappa g \, ,
\end{equation}
This relation is a direct result of the fact that the phonon scattering rate of $\tau$-TLSs is $\propto \gamma_{\tau}^2$, but their DOS is $\propto 1/\gamma_S \gamma_{\tau}$, as it is dictated by the coupling to the $S$-TLSs (or, in the absence of $S$-TLSs, to the strong disorder, see discussion of mixed crystals in Sec.~\ref{Further}).

Assuming the usual value $\kappa \approx 0.1$,
we then find that $C_o \approx 10^{-3}$. This small and universal value for $C_o$ derives from
its dependence on $g$, which is also small and
varies little between different strongly disordered materials. Notice that $g$ also
dictates the universal crossover energy $T_U \approx 0.2 g J_o$ consistent with the
observed energy scale of $\approx 3$K.
Although glassiness and the mechanism leading to it are not required by
our theory, for glassy systems $T_G \approx J_o$, i.e. $T_U \approx 0.2 gT_G$, $\gamma_{\tau}^2
\propto T_{\rm G}$, and $P_o
\propto T_{\rm G}^{-1}$, in agreement with experiments\cite{RP80,DMS83}.

\subsection{Further experimental consequences}
\label{Further}

{\it Strong dilution} :
The low temperature universal phenomena are observed in both amorphous solids and disordered
lattices. One advantage of the latter is the ability to control the identity of the host ions and the concentrations of the tunneling impurities. For example, in KBr:CN, experiments find that
phenomena such as the temperature independence of the internal friction and $T^2$ dependence of
the thermal conductivity exist below $3$K for CN concentrations $0.2<x<0.7$, but not for $x<0.2$.
However, once CN impurities are added to the mixed crystal KBr$_{0.5}$KCl$_{0.5}$, then the above mentioned phenomena are observed also at $x \ll 1$, only with a tunneling strength which is
proportional to $x$. These experimental results \cite{watson95,pohlT99} were argued to support the notion that the loss of universality at $x \ll 1$ is a result of the reduction of the strain. Our analysis supports this notion.

Let us consider first the dilute case in a pure crystal, e.g. KBr$_{1-x}$CN$_x$ with $x \ll 1$.
In this system the strain at each CN impurity site is caused solely by the presence of the other
CN impurities. This does not change the typical value of $\gamma_S$, but $\gamma_{\tau} \sim
x^{4/3}$, see Eq.(\ref{gammainvest}). As a result, the typical energy for an S excitation $J^{SS}
\propto x$, whereas the typical energy for a $\tau$ excitation $J^{S\tau} \propto x^{7/3}$.
Furthermore, the magnitude of the $\tau$-TLS energy at site $j$ and its coupling to the phonon
field are strongly correlated, as both are dictated by the proximity of the nearest neighbor
impurity.

For strongly disordered systems we have argued that universal properties appear below $T_U$ since at this energy scale $\tau$-TLSs behave as noninteracting TLSs with a roughly constant DOS, and with an interaction
with the phonon field which is not correlated with their energy. At $x \ll 1$ $T_U(x) \propto
x^{7/3} \ll 3$K. Furthermore, even at $T < T_U(x)$ it is not clear that universality exists, as
it may well be violated by the above mentioned correlations.

Let us now consider the case of dilute tunneling impurities in a mixed crystal, as were realized
by CN impurities added to KBr$_{0.5}$Cl$_{0.5}$\cite{watson95,pohlT99}. In this case the strain
at the CN impurity sites is dictated by the mixed lattice itself, and is comparable to the strain
in KBr$_{1-x}$:CN$_x$ with $0.2<x<0.7$. The typical $S$ and $\tau$ energies are now dictated by
the random field terms $h^S$, $h^{\tau}$ arising from the interaction of the CNs with the Br and
Cl ions through the volume term in Eq.(\ref{H-full}) (or its simplified form, Eq.~(\ref{V-cont-disorder}) in App.~\ref{SEC:TLSPhonon}), see also Ref.\cite{MS08}. $\gamma_{\tau}$ can be derived similarly to the derivation in Sec.\ref{SEC:TLSPhonon}. Thus, all energy scales are the
same as in the strongly disordered KBr:CN. The only effect of having small $x$ is the reduction
in $n(\tau)$, and we therefore find that universal properties exist in the mixed crystals, albeit with a tunneling strength $C \approx 0.1 g x/x_c$. This result is in agreement with
experiments \cite{watson95,pohlT99}.

{\it Electric dipole interactions}: Both $\tau$ and $S$ defects
may have electric dipole moments, leading to electric dipole
mediated interactions whose strength at the nearest impurity distance is
$J_{ee}$. Thus, $J_o > J_{ee}$ almost always, but the ratio
$J_{ee}/J_o^{S \tau} = J_{ee}/gJ_o$ can be small or large, and
typically $J_{ee} > J_o^{\tau \tau} = g^2 J_o$. If $J_{ee} < g J_o$,
then all the arguments above go through, but now the $\tau - \tau$
interaction strength is $J_{ee}$; the energy at which the very weak
dipole 'depression' appears in $n_{\tau}(E)$ then becomes $J_{ee}$.
Such dipole depressions in the DOS are seen in some
experiments \cite{osheroff} below $1$K.
If $J_{ee} > g J_o$, electric dipole interactions also change the
$S-\tau$ interactions. This does not change the basic picture, but
now one finds $T_U \sim J_{ee}$, and $C_o$ is reduced to $C_o \sim
g^2 J_o/J_{ee}$.

{\it Amorphous systems and glasses}: Experimentally, there is
overwhelming support to the notion that the universal phonon attenuation
at low temperatures is equivalent in disordered lattices and amorphous
solids. This strongly suggests that the mechanism leading to universality is
intrinsic to the disordered state of matter, and is the same in both systems.
One can therefore expect that one model would be relevant to both disordered lattices
and amorphous solids. Our model is derived microscopically for the
disordered lattices. We argue here that although amorphous solids lack long
range order, and exact symmetries on any scale,
it is plausible that our model describes well their low temperature
characteristics.

While amorphous materials
do not show long range order, local order with a lattice bond
coordination number does exist, except at defect sites, and
inter-site distances are hardly altered from lattice values.
For example, in amorphous Si, $\delta a/a_o \sim 0.03 (0.1)$
for nearest (next nearest) neighbors \cite{TB12},
and the nearest-neighbor bond angle differs by a maximum of $10\%$. Now,
our results depend essentially on the distinction between $S$ and
$\tau$ defects and rely on local properties of the system (contributions
to $\gamma_{\tau}$, resulting from deviations from inversion symmetry,
are random and decrease as $1/r^4$, see App.~\ref{SEC:TLSPhonon}). Thus, neither $J_o$ nor
$g$ is strongly affected by amorphicity.

This allows us to make some clear predictions for amorphous systems.
First, if nearly inversion-symmetric TLSs
do exist therein, then our model predicts that amorphous
systems will also show universality with similar values of $T_U$ and
$C_o$. Second, we predict that in addition to the TLSs responsible for the universal properties,
there exists a second type of
TLS with much higher (by a factor $\sim \cdot 10^2$) coupling to phonons and with a DOS given in Fig. 2. This prediction could in principle be
checked by using the powerful technique of phonon echo \cite{GG76,GHSD79,NFHE04}, adjusted to fit
the above characteristics of the asymmetric excitations.

\section{Conclusions}
\label{sec:conclusions}

We find that the low temperature universal properties in disordered solids
result from inversion symmetric TLSs. Such TLSs interact weakly with phonons, yet gap other
non-symmetric TLSs at energies lower than $3$K. Quantitative universality and the energy scale of 3K below which universality is observed are both dictated by the rather universal value of the
ratio of strain to interatomic distance in strongly disordered solids. This is because this value dictates both the relation between the DOS of the TLSs and their coupling to the phonon field, and the ratio between the universality temperature and the glass transition temperature. Various additional experimental observations, some of which are long unaccounted for, are naturally explained within our theory. Our results are derived
from the microscopic properties of disordered lattices, and their applicability to amorphous
solids needs to be checked.

{\it Acknowledgments} ---
We thank A. Aharony, A. Burin, O. Entin-Wohlman, A. Gaita-Ari\~no, and A. J. Leggett for
discussions. The work was supported
by NSERC in Canada, PITP, and the ISF.

\appendix

\section{Derivation of Model, and mapping to interacting $S$ and $\tau$-spins}
\label{Sec:Mapping}

In the main text we use a model of interacting $S$ and $\tau$ spins to describe the
low-energy excitations in the system. In this section we give more details of how this model is
derived, starting from a microscopic model of the systems we are
interested in.

\vspace{3mm}

\subsection{Microscopic Model: Single Impurity}
\label{microM}

Consider first a single impurity or defect sitting in an otherwise perfect lattice, on site $j$
at position ${\bf r}_j$. We assume the lattice itself is inversion-symmetric, and that the system is insulating. We model the system by a Hamiltonian $\hat{\cal H} = \hat{H}_{latt} +
\hat{H}_{imp} + \hat{V}$, where $H_{latt}$ is the bare lattice Hamiltonian, $\hat{H}_{imp}$ is
the impurity Hamiltonian, and $\hat{V}$ is their mutual interaction. One can then write the
impurity Hamiltonian as $\hat{H}_{imp} = \hat{H}^o_{imp} + \hat{H}^T_{imp}$, where the first bare
or 'potential' term takes the form
\begin{equation}
\hat{H}^o_{imp}({\bf r}_j) \;=\; \sum_{n=1}^N \sum_{\sigma}^{\pm} \epsilon_{jn} c^{\dagger}_{j n
\sigma} c_{j n \sigma}
 \label{H-o}
\end{equation}
and the second kinetic or 'tunneling' term is written as
\begin{equation}
\hat{H}^T_{imp}({\bf r}_j) = \sum_{nn'} \sum_{\sigma \sigma'} t_{nn',\sigma \sigma'}
c^{\dagger}_{j n \sigma} c_{j n' \sigma'}
 \label{H-T}
\end{equation}
We will drop this second tunneling term, for reasons explained below.

In these equations, $c^{\dagger}_{j n \sigma}, c_{j n \sigma}$ create/destroy an impurity or
defect state labeled by (i) the impurity site $j$, (ii) the pair index $n$ for the $N$ different
excitation pairs on the $j$-th site, and (iii) the internal pair quantum number $\sigma = \pm$. Thus we
are assuming a set of $N$ pairs (often called 'two-level systems, or TLS's) of 'internal states'
for the impurity/defect, to give a total of $2N$ states. That they come in degenerate pairs, with
energy $\epsilon_{jn}$, follows from the assumed inversion symmetry. The number and physical
nature of these states depends on the system of interest. Thus, eg., a light Halide defect like a
$Li$ atom, which substitutes for $Na$ or $K$ in a $KCl$ or $NaCl$ lattice, can rattle around
inside a 'cage' formed by the original lattice. The 8 lowest energy  $Li$ impurity states
comprise 4 degenerate pairs of states; each pair is degenerate with the other 3 pairs, and each state is quasi-localized around potential minima at the $<111>$
sites. The same is found if we substitute in $CN$ impurities. Those states quasi-localized around other 'cage' sites (eg., the $<100>$ sites) are higher in energy. However, if we substitute in
$OH^-$ impurities, we find 6 degenerate lowest states localized around 3 pairs of minima at the
$<100>$ sites; whereas if we substitute in $F$ or $Ag$ impurities into a $NaBr$ lattice, we find
6 pairs of degenerate low-energy states localized around the $<110>$ sites. In all these cases,
the set of energies $\epsilon_{jn}$ divides into several degenerate groups, one of which is
lowest in energy. Typically each such group of degenerate states contains more than one pair of
states, simply because there are other symmetries at the impurity site apart from inversion
symmetry. The number of impurity states that we consider in our Hamiltonian depends on the UV
cutoff we assume for the impurity effective Hamiltonian. Thus, in the case of, say, $CN$
impurities in a  $KBr$ or $KCl$ lattice, we might want to consider both the lowest set of 4 pairs at the $<111>$ sites, and the set of 3 pairs of $<100>$ sites which are at somewhat higher
energy. This then makes for a total of 14 states in our impurity Hilbert space.

In more general cases the physical locations of the TLS states may not be obvious - the
inversion-symmetric pairs of states may refer more complicated defects, or to some kind of
rotation. Finally, whereas for systems with only one kind of defect the energy $\epsilon_{jn}$ is independent of site index $j$, we can also consider a lattice having several different kinds of
impurity or defect at different sites. The same model Hamiltonian still applies; but the energy
$\epsilon_{jn}$ will then depend on $j$.

As a simple example of this physics we consider a 4-site 'toy' model. This model was already introduced in Fig. 1(a) of the
main paper. It has the site Hamiltonian
\begin{equation}
\hat{H}^o_{imp} \;=\; \sum_{\mu = 1}^4 \epsilon_{\mu} c^{\dagger}_{\mu} c_{\mu}
 \label{HoToy1}
\end{equation}
in which, because of inversion symmetry, $\epsilon_1 = \epsilon_2 = \epsilon_A$, and $\epsilon_3
= \epsilon_4 = \epsilon_B$, where $A,B$ label the two different inversion-symmetric pairs. We can
thus also write this Hamiltonian as
\begin{eqnarray}
\hat{H}^o_{imp} &=& \sum_{\sigma} \epsilon_A c^{\dagger}_{A \sigma} c_{A \sigma} + \epsilon_B
c^{\dagger}_{B \sigma} c_{B \sigma} \nonumber \\
&=& E^o_{AB} + \Delta^o_{AB} \sum_{\sigma} (\epsilon_A c^{\dagger}_{A \sigma} c_{A \sigma} -
\epsilon_B c^{\dagger}_{B \sigma} c_{B \sigma})
 \label{HoToy2}
\end{eqnarray}
where $E^o_{AB}$ is the mean or 'midpoint' energy of the $A$ and $B$ states, and $\Delta^o_{AB}$
is the difference in energy between them.

We have ignored any tunneling between the different states in this toy model, just as we did in
(\ref{H-T}), because the tunneling amplitudes - typically $|t_{nn',\sigma \sigma'}| \sim O(1~K)$ or less, are small, in strong disorder, compared with the bias energies of the $S$ TLSs - typically $\approx 500$K and the bias energies of the $\tau$ TLSs, typically $\approx 10$K, as is discussed in Sec. \ref{SEC:TLSPhonon}. Thus, the effect of tunneling on the distribution of biases of the symmetric and asymmetric TLSs is negligible. Once the latter is established, tunneling resumes a role identical to its role in the ST model.
The differences $ \Delta \epsilon_{nn'} = |\epsilon_{jn} - \epsilon_{jn'}|$ will vary greatly between different systems, but as discussed below, we will only be interested in states for which $\Delta \epsilon_{nn'}$ is less than a few hundred Kelvin, since only such states can be shifted by the strain to low energies.

Consider now the defect-phonon interaction, which arises because the defect distorts the lattice
(since the system is neutral, we ignore electron transfer terms). It is helpful to work
this out first for the toy model. We have an interaction
\begin{equation}
\hat{V}_j \;=\; \sum_{\mu =1}^4 V_{j\mu}(u_{\alpha \beta}) c^{\dagger}_{j\mu} c_{j\mu}
 \label{Vsite}
\end{equation}
at the j-th site, where $V_{j\mu}(u_{\alpha \beta})$ is some function of the lattice strain tensor $u_{\alpha \beta} ({\bf r}_j)$ at site $j$. We now introduce the Ising variables $\hat{S}^z$ and $\hat{\tau}^z$ as follows (compare main text, Fig. 1(b)). Define (i) the change $\bar{\eta}_{AB} = E_{AB} - E^o_{AB}$ in the midpoint energy of the 4 states caused by the defect-phonon coupling; (ii) the change $\bar{\Gamma}_{AB} = \Delta_{AB} - \Delta^o_{AB}$ in the splitting between the $A$ and $B$ states, and (iii) the splitting energy $\bar{\zeta}_{n} = \epsilon_{n \uparrow} - \epsilon_{n \downarrow}$, opened up by defect-phonon coupling, between the internal states of the $n$-th pair (here $n = A,B$).

In terms of phonon variables, $\bar{\eta}_{AB} = \eta_{AB} \delta^{\alpha \beta} u_{\alpha
\beta}$, where $\eta_{AB}$ is just an isotropic energy shift, and $\bar{\Gamma}_{AB} =
\Gamma_{AB}^{\alpha \beta} u_{\alpha \beta}$; the lattice strain is sensitive to the difference between the $A$ and $B$ states because these are not related to each other by inversion. However, the 2 internal states of a given pair {\it are} related by inversion, and so only the gradient $\partial_{\gamma} u_{\alpha \beta}$ of the lattice strain can cause a splitting between them (see Fig. {\ref{secondderivative}). The splitting energy is $\bar{\zeta}_{n} = \zeta_n^{\alpha \beta \gamma} \partial_{\gamma} u_{\alpha \beta}$.

\begin{figure}
\begin{center}
\hspace{-1cm}
\includegraphics[width = 0.8\columnwidth]{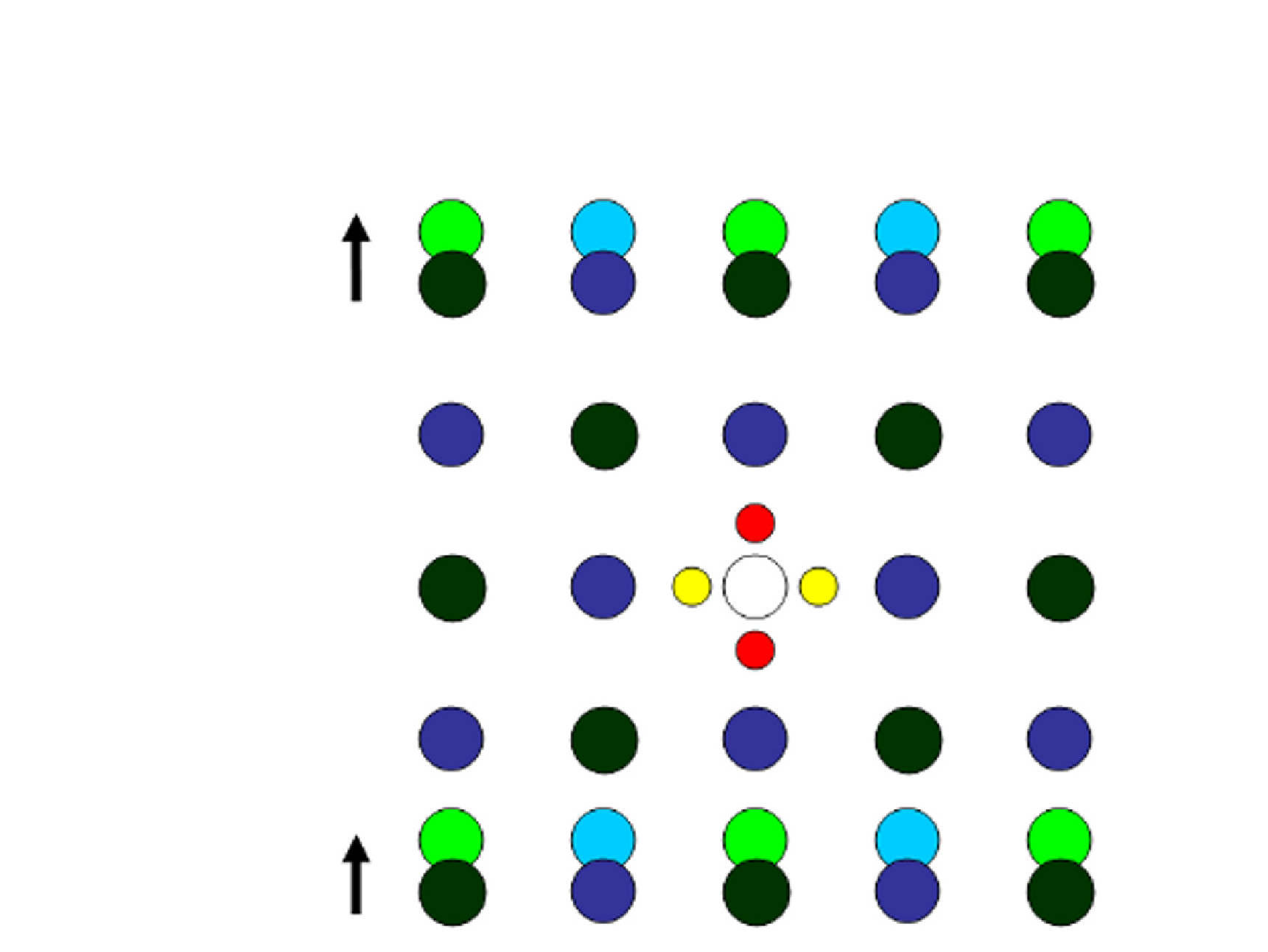}
\end{center}
    \caption{Interaction of the symmetric TLSs with the gradient of the strain. The
    degeneracy of states related by inversion symmetry is broken by the second
    derivative of the phonon displacement. }
    \label{secondderivative}
\end{figure}

Thus we see that the interaction for the toy model, with an impurity at the $j$-th site, is (here we use the Einstein summation convention for spin indices $\alpha, \beta$, etc.):
\begin{eqnarray}
\hat{V}_j \;=\;  [ \eta_{AB} \delta^{\alpha \beta} &+&  \Gamma_{AB}^{\alpha \beta} \hat{S}_{AB}^z
] u_{\alpha \beta}({\bf r}_j) \nonumber \\ &+& \sum_n^{A,B} \zeta_n^{\alpha \beta \gamma}
\hat{\tau}_{n}^z \;  \partial_{\gamma} u_{\alpha \beta}({\bf r}_j)
 \label{V-toy}
\end{eqnarray}
We have defined $\hat{S}_{AB}^z = (|A \rangle \langle A| - |B \rangle
\langle B |)$, and likewise $\hat{\tau}_n^z = (|n \uparrow \rangle \langle n \uparrow | - |n
\downarrow \rangle \langle n \downarrow |)$, in terms of the site operators given above. We emphasize here that (i) we should {\it not} think
of $\hat{S}_z$ as the $z$-component of a spin variable - we are treating it purely as an Ising
variable; and (ii) we can if we wish define a spin-half operator $\hat{\bm{\tau}}_n$, for a
specific pair $n$, with the transverse terms $\hat{\tau}^{\pm}_n$ describing tunneling processes
between internal states of a given pair. However, since we ignore such tunneling processes in
this paper, we only consider $\hat{\tau}_n^z$, and treat it as an Ising variable as well.

Generalizing to an arbitrary number of levels on the impurity site we get the final Hamiltonian
for a lattice system with a single impurity at site $j$:
\begin{widetext}
\begin{equation}
\hat{H}_{imp} \;=\; \hat{H}_{ph} + \hat{H}^o_{imp}({\bf r}_j) \;+\; \sum_{nn'} \sum_{\alpha
\beta} \;\{ [ \eta_{jnn'} \delta^{\alpha \beta} +  \Gamma_{jnn'}^{\alpha \beta} \hat{S}_{jnn'}^z]
u_{\alpha \beta}({\bf r}_j) + \delta_{nn'} \zeta_{jn}^{\alpha \beta\gamma} \hat{\tau}_{jn}^z \;
\partial_{\gamma} u_{\alpha \beta}({\bf r}_j) \}
 \label{H-full}
\end{equation}
\end{widetext}
where $\hat{H}^o_{imp}({\bf r}_j)$ is just the bare impurity term in (\ref{H-o}), and
$\hat{H}_{ph}$ is the acoustic phonon Hamiltonian, which we write as
\begin{equation}
\hat{H}_{ph} = \sum_j  C_{\alpha \beta \gamma
\delta}({\bf r}_j) u_j^{\alpha \beta} u_j^{\gamma \delta}
 \label{Hph}
\end{equation}
where the form of $C_{\alpha \beta \gamma\delta}$ depends on the symmetry of the lattice.

\subsection{TLS-phonon interactions and the random strain field}
\label{SEC:TLSPhonon}

We now consider a system with a finite concentration of impurities/defects at random sites $\{
{\bf r}_j \}$.
The key point we address here is the way in which the random strain fields in the system then
modify the interactions between the defects, and generate a new effective Hamiltonian (the one
given in Eq.(1) of the main text).

The phonon-generated interaction between a pair of defects has been discussed in many papers
\cite{halp77,MS08}. The strength of the interactions $J^{ab}$ (where $a, b$ label either $S$ or $\tau$) can be found microscopically from the direct phonon exchange between them
\cite{halp77,MS08}, 2nd-order in the defect-phonon interaction. One finds, for a pair of
defects at sites $i$ and $j$ in an otherwise perfect crystal, that
\begin{equation}
J_{ij}^{ab} \sim \frac{ f(\theta_{ij}) {\Gamma}_a {\Gamma}_b}{\rho c^2 R_{ij}^{3+\alpha}}
 \label{JJ-mag}
\end{equation}
where $R_{ij}$ is the inter-defect distance, and this formula is only valid for distances $\gg
a_o$, the lattice parameter. All of the very complicated angular dependence of these interactions \cite{halp77,MS08} is incorporated into the angular factor $f(\theta_{ij})$, where
$\theta_{ij}$ is the polar angle between the defects. The parameter $\alpha = 0,1,2,$ for $\{ab\} = \{SS\}, \{S\tau\}$, and $\{\tau,\tau\}$ respectively; and we write $\Gamma_S = \Gamma$ and
$\Gamma_{\tau} = \zeta$, where $\Gamma = |\Gamma_{nn'}^{\alpha \beta}|$ and $\zeta =
|\zeta_n^{\alpha \beta \gamma}|$, and we have suppressed the pair indices in these constants
because we won't need them in what follows.  We note that $J^{S\tau}_{ij}$ and
$J^{\tau\tau}_{ij}$ fall off like $R_{ij}^{-4}$ and $R_{ij}^{-5}$ respectively, and so at long
ranges can be neglected compared to $J_{ij}^{SS} \sim R_{ij}^{-3}$. The more rapid fall-off
occurs because the phonons couple to the $\tau$-defects via the gradient of the phonon strain
rather than the strain itself, adding extra powers of $1/R_{ij}$ to the interaction
\cite{MS08}.

However, when we have a finite random concentration of defects, things change in two important
ways, viz. (i) a set of random strain fields is generated \cite{MS08} in the system, which has
the effect of breaking the inversion symmetry randomly at each site - this radically modifies the coupling of the $\tau$-spins to the phonons; and (ii) because of this, the strength of the
interactions $J_{ij}^{S\tau}$ and $J_{ij}^{\tau\tau}$ is changed, and this is crucial for the
universality.

To deal with this physics is quite subtle. In ref. \cite{MS08}, this was done by simply summing
independently the random strain fields generated by each impurity. However this is not quite
correct, because as the strain fields increase in strength, they alter the defect-phonon coupling to the $\tau$-defects themselves. Thus the results in ref. \cite{MS08} are valid only in the
regime where the concentration of defects $x \ll 1$, whereas here we are interested in strong
disorder, where $x$ is not small, and a new calculation is required.

The key to the physics in this strong disorder regime is to realize that we are dealing with
a 3-body effect - we must recalculate the 2-defect interaction (\ref{JJ-mag}) in the presence of
a 3rd defect \cite{noteSasha}. Without this 3rd defect, as we have seen, the effective $\tau_i^z S_j^z$ interaction is $\propto 1/R^4$.  We thus recalculate the effective $\tau_i^z S_j^z$ interaction, but now in third order in the defect-phonon coupling. We will also assume that the 3rd impurity $k$ is close to
impurity $i$. This is because (a) the $1/R_{ik}^4$ dependence of the interaction, and (b) the dependence of its sign, on the angle between sites $i$ and $k$, means that such close '3rd party' configurations will dominate all sums over $k$, and the deviation from inversion symmetry at the $\tau$ impurity.

We start again from the bare defect-phonon coupling in (\ref{H-full}), now written as a sum over
defects, so that we have
\begin{equation}
\hat{V} \;=\; \sum_{\{j \}} \left[  \eta_j
u_j^{\alpha \alpha} + \Gamma^{\alpha
\beta}_j u_j^{\alpha \beta} S_j^z
+ \zeta^{\alpha \beta \gamma}_j
\partial_{\gamma} u_j^{\alpha \beta} \tau_j^z  \right].
 \label{V-cont}
\end{equation}
where we have dropped the pair indices $n,n'$ since they play no role in what follows.
We write the Fourier transform of the lattice displacement field as
\begin{equation}
X_{\alpha}(x) = \frac{1}{\sqrt{N}} \sum_{q,\mu} X_{q \mu} {\bf
e}_{q \mu, \alpha} e^{i q x} \label{Fourier}
\end{equation}
where ${\bf e}_{q, \mu, \alpha}$ is a phonon polarization index. We
then minimise the total potential energy, i.e. the sum of the bare
phonon potential energy plus the interaction term above, to find the
resulting distortion in the lattice\cite{MS08}. This distortion is then found
to be
\begin{equation}
\delta X_{\alpha} (x) = \frac{1}{2} \sum_{q \mu} (\delta X_{q \mu}
{\bf e}_{q \mu \alpha} e^{i q x} + H.c.),
\end{equation}
where we have
\begin{widetext}
\begin{equation}
\delta X_{q\mu} = \frac{1}{\sqrt{N} M \omega_{q \mu}^2} \left(
\sum_{\gamma,\delta,\eta} \zeta_i^{\gamma \delta \eta} {\bf e}_{q
\mu \gamma} q_{\delta} q_{\eta} e^{-i q x_i} \tau_i^z + i
\sum_{\alpha \beta} \gamma_j^{\alpha \beta} {\bf e}_{q \mu \alpha}
q_{\beta} e^{-i q x_j} S_j^z \right) .
\end{equation}
and we have restored the sums over spin indices to make clearer what is being summed.

The effect of this distortion on the 3rd impurity $k$ comes only
from the phonon term (\ref{Hph}), which we write in terms of the lattice displacement as
\begin{equation}
\hat{H}_{ph} \;=\; \sum_{\rho \phi \chi \psi} C_{\rho \phi \chi \psi}({\bf
r}_k) \frac{\partial \delta X_{k \rho}}{\partial x_{k \phi}}
\frac{\partial \delta X_{k \chi}}{\partial x_{k \psi}}
\label{elastictensor}
\end{equation}
where
\begin{equation}
\frac{\partial \delta X_{k \rho}}{\partial x_{k \phi}} = \sum_{q
\mu} \frac{1}{NM \omega_{q \mu}^2} \left( \sum_{\gamma \delta \eta}
\zeta_i^{\gamma \delta \eta} {\bf e}_{q \mu \gamma} q_{\delta}
q_{\eta} \sin{q (x_k-x_i)} \tau_i^z + \sum_{\alpha \beta}
\gamma_j^{\alpha \beta} {\bf e}_{q \mu \alpha} q_{\beta} \cos{q
(x_k-x_j)} S_j^z \right) {\bf e}_{q \mu \rho} q_{\phi} .
\end{equation}
We now evaluate the term in Eq.(\ref{elastictensor}) proportional to
$\tau_i^z S_j^z$, in order to find the change $\delta J^{S \tau}_{ij}(k)$
in the $\tau_i^z S_j^z$
interaction caused by the impurity at site $k$ . Defining
\begin{equation}
A_{q \mu} \equiv \sum_{\gamma \delta \eta} \zeta_i^{\gamma \delta
\eta} {\bf e}_{q \mu \gamma} q_{\delta} q_{\eta} \tau_i^z
\end{equation}
\begin{equation}
D_{q \mu} \equiv \sum_{\alpha \beta} \Gamma_j^{\alpha \beta} {\bf e}_{q \mu
\alpha} q_{\beta} S_j^z ,
\end{equation}
we find the shift in the $S\tau$ interaction caused by the
impurity at site $k$ to be
\begin{equation}
\delta J^{S \tau}_{ij}(k) = 2 \sum_{\rho \phi \chi \psi} C_{\rho \phi \chi
\psi}({\bf r}_k) \sum_{q \mu} \frac{{\bf e}_{q \mu \rho} q_{\phi}}{N
M \omega_{q \mu}^2} A_{q \mu} \sin{[q (x_k-x_i)]} \sum_{q' \mu'}
\frac{{\bf e}_{q' \mu' \chi} q'_{\psi}}{N M \omega_{q' \mu'}^2}
D_{q' \mu'} \cos{[q' (x_k-x_j)]} . \label{Ekijfull}
\end{equation}
\end{widetext}
In the acoustic approximation for the phonon spectrum we get, after similar integrations to those discussed in ref. \cite{MS08}, the result
\begin{equation}
\delta J^{S \tau}_{ij}(k) = \frac{c_{ij} \zeta_i \Gamma_j C_k}{\rho^2 c^4 R_{ik}^4
R_{jk}^3}
 \label{delJ}
\end{equation}
where $\zeta_i = |\zeta_i^{\alpha \beta \gamma}|$, and similarly for $\Gamma_j, C_k$, and all
angular dependence has been absorbed in $c_{ij} \sim O(1)$. This is a complicated function of
angle, which depends on the position of the impurities and takes either sign. It can then be
treated as a random variable.

Actually here the acoustic approximation is not valid, because it is easily seen, as stated
above, that when we sum over all the different impurities at positions ${\bf r}_k$ to find the
total change $\Delta J^{S\tau}_{ij}$ in $J^{S\tau}_{ij}$, the dominant configurations will have
$R_{ik} \approx a_o$. However, since we are only interested in an estimate, we make the following
assumptions:

(i) we assume a defect concentration $x$ which is not small, which means that the probability
that a defect, at some site $i$, will have a "3rd-party" defect on a neighbouring site $k$ is $\propto x^{1/3}$.

(ii) this implies that $R_{ik} \sim a_o x^{-1/3} \ll R_{jk}$, and so we find

\begin{equation}
\Delta J^{S \tau}_{ij} \approx \frac{\gamma_S \gamma_{\tau}}{ \rho c^2 R_{ij}^3}
\end{equation}
where
\begin{equation}
\gamma_{\tau} \approx \frac{\bar{\zeta} \bar{C} x^{4/3}}{\rho c^2 a_o^4 } .
\label{gammainvest}
\end{equation}
and where

\begin{equation}
\gamma_S \approx \bar{\Gamma} .
\label{GammaSinvest}
\end{equation}
Here overbars on the various quantities indicate typical values.

Let us now estimate the magnitude of $\gamma_{\tau}$. First, $\bar{C} \cdot (\delta a/a_o)^2
\approx \delta M \omega^2 (\delta a)^2$, where $\delta a$ denotes a typical strain, and
the latter expression is the difference in kinetic energy of the impurity compared to
the host ion. Approximating $\delta M \approx M$ and $\omega \approx c/a_o$ we obtain
$C \approx M c^2$, and thus we get $\gamma_{\tau} \approx x^{4/3}
\zeta/a_o$.

Let us now estimate $\bar{\zeta}$. Since $\zeta$ is the coefficient
of the second derivative of the displacement
(see Fig. \ref{secondderivative}), we have $\bar{\zeta} \cdot \delta a/a_o^2 \approx E_C \cdot (\delta a/a_o)^2$,
where $E_C$ is the typical (Coulombic) electronic energy involved in charge displacement
in the solid. Similarly, $\bar{\Gamma}\delta a/a_o \approx E_C \delta a/a_o$, i.e.
$\gamma_S \approx E_C$.
Thus, for $x \approx 1$ we find $\gamma_{\tau} \approx \gamma_S \delta a/a_o$.
The parameter $g \equiv \gamma_{\tau}/\gamma_S \approx \delta a/a_o$ is the small parameter of
our model. Since $E_{\rm \Phi}/E_C \approx \delta a/a_o$, where
$E_{\Phi}$ is the characteristic energy of elastic deformations in the solid
(ie., roughly the Debye energy), we also have $g \sim E_{\Phi}/E_C$.

\vspace{2mm}

Thus, in strongly disordered systems, deviations from inversion symmetry result in a finite interaction between the $\tau$-TLSs and the phonon field, and the TLS-phonon interaction Hamiltonian is changed from Eq.(\ref{V-cont}) to

\begin{equation}
\hat{V}_{dis} \;=\; \sum_{\{j \}} \left[  \eta_j
u_j^{\alpha \alpha} + \Gamma^{\alpha
\beta}_j u_j^{\alpha \beta} S_j^z
+  \gamma^{\alpha
\beta}_j u_j^{\alpha \beta} \tau_j^z  \right].
 \label{V-cont-disorder}
\end{equation}
with $|\gamma^{\alpha \beta}| \equiv \gamma_{\tau}$, and $\gamma_{\tau} \approx g \gamma_S$. This change in the form of the interaction Hamiltonian upon the introduction of disorder has been recently confirmed by molecular static calculations\cite{CBS13}. Starting from the interaction Hamiltonian in Eq.(\ref{V-cont-disorder}), one readily obtains the effective Hamiltonian for the system of the form

\begin{eqnarray}
\hspace{-0.5cm}
H_{S \tau} &=& \sum_j [h_j^S S_j^z + h_j^{\tau} \tau_j^z] \nonumber \\
&+& \sum_{ij} [J_{ij}^{SS} S_i^z S_j^z  +
J_{ij}^{S\tau} S_i^z \tau_j^z + J_{ij}^{\tau \tau} \tau_i^z
\tau_j^z]
 \label{H-StApp}
\end{eqnarray}
where the interaction strengths $h_{a}, J_{ab}$, are now random variables.
If we write the size
of the random couplings in terms of $R_o \sim a_o x^{-1/3}$, the mean distance between the
defects, one has
\begin{equation}
J_{ij}^{ab} = \frac{c_{ij}^{ab} {\gamma}_a {\gamma}_b}{\rho c^2 (R_{ij}^3 + \tilde{a}^3)} \equiv
J^{ab} {R_o^3 \over R_{ij}^3 + \tilde{a}^3}
 \label{hJ-magApp}
\end{equation}
where again $c_{ij}^{ab} \sim O(1)$, where we introduce $\tilde{a}$ as a short distance
cutoff for the interaction, and where
\begin{eqnarray}
J^{SS} \sim J_o &\equiv& {\gamma_S^2 \over \rho c^2} \sim 500~K \nonumber \\
J^{S\tau} \sim gJ_o \sim 10~K \;\;&;&\;\; J^{\tau\tau} \sim g^2 J_o \sim 0.2~K
 \label{Js}
\end{eqnarray}
The numerical values for the energies assume typical values for the parameters in disordered
insulators.
The random fields are governed by the same parameters; their typical strengths are $h^S \lesssim
J_o$ and $h^{\tau} \lesssim g J_o$, as was found earlier \cite{MS08}.

The value of $J_o$ denotes the typical energy for an asymmetric excitation of a single impurity in the static lattice, whereas $g J_o$ denotes the typical energy for a symmetric excitation. Thus, the energy spectrum of a single impurity is as shown in Fig.1(c) in the main text. Our approximation here of suppressing the indices $n, n`$ is equivalent to considering for each impurity the lowest two pairs of levels, which is justified at low temperatures.

The Hamiltonian (\ref{H-StApp}) is the one quoted in Eq. (1) of the main text. Its form was already discussed in Ref. \cite{MS08}; what is new here is the calculation of the renormalization of $J^{S\tau}_{ij}$ by the disorder, using the 3-body technique described above.

\section{Densities of states for the interacting $S$-$\tau$ system}
\label{SEC:gapdos}

As discussed in the text, the interaction between the $S$ and $\tau$ pseudospins has a drastic
effect on the densities of states of the excitations in the system. What we wish to do here is
find the densities of states (DOS) for the $S$ and $\tau$ excitations.

\begin{figure*}[ht!]
\centering
\subfigure[]{\includegraphics[scale=0.18]{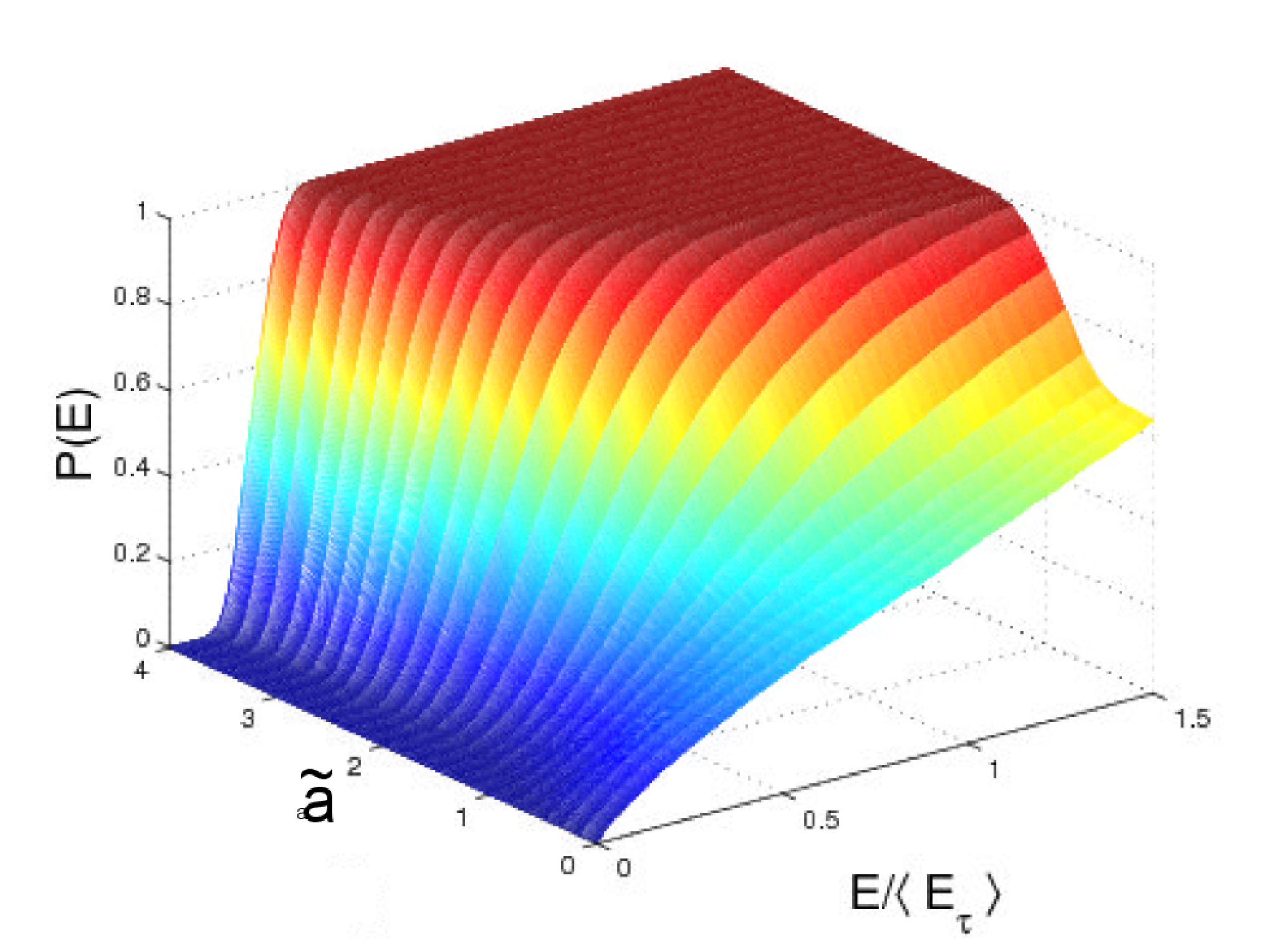}}
\subfigure[]{\includegraphics[scale=0.36]{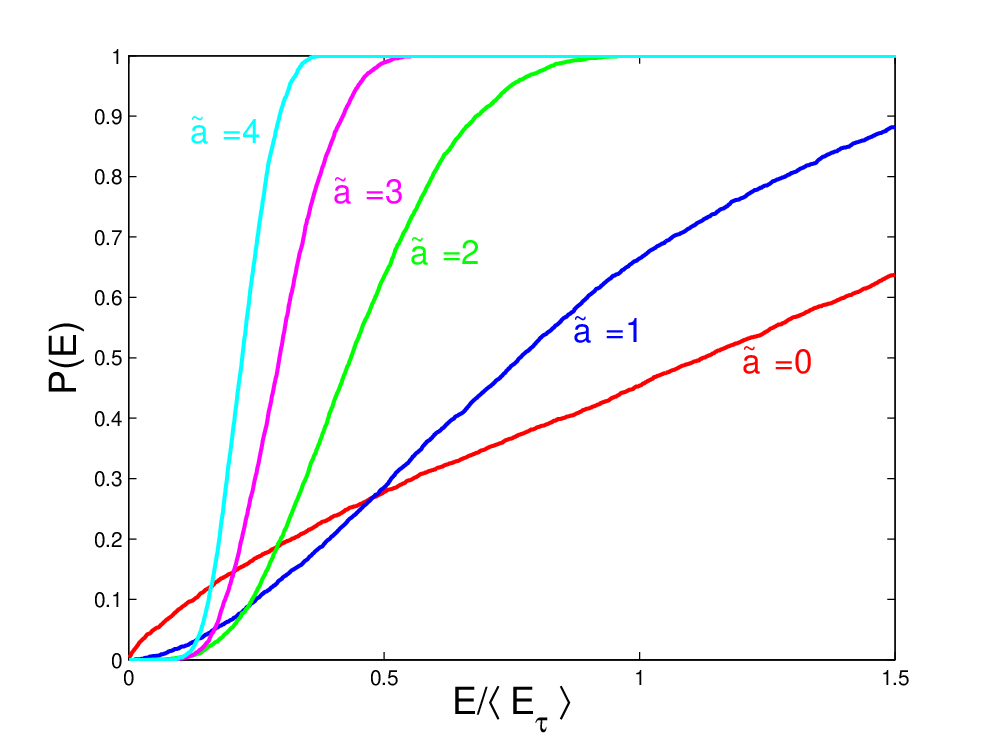}}
\caption{ The true $S$-spin density of states $n_S(E)$, as a
function of energy for different short-distance behaviors of the
interaction. Details are as in Figure 2 in the main text, only here x$=0.5$ and lattice
size is $13^3$
cells (ie., $\sim 4400$ TLSs).}
\label{dos05}
\end{figure*}

\subsection{DOS of $\tau$ impurities}

Let us start from the effective Hamiltonian in (\ref{H-StApp}), but for the moment ignore the random
fields, and just work with the interaction terms. We begin with the mean field DOS introduced in
the main text, viz., the $S$-defect DOS
$n^o_S(E)$ and the $\tau$-defect DOS $n^o_{\tau}(E)$, having widths $J_o, gJ_o$ and peak heights
$\propto 1/J_o,
1/gJ_o$, respectively. Now we consider the effect on these two DOSs of the $S\tau$ interaction
$J^{S\tau}$.

Consider first the effect exerted by the S spins on the $\tau$ spins,
neglecting the very small $\tau-\tau$ interaction $J^{\tau\tau}$. Following Efros
and Shklovskii \cite{efros75}, we can argue that the new DOS is given by
\begin{eqnarray}
n_{\tau}(E) &=& n^o_{\tau}(E) \prod_j \Theta(E + E_{S_j} -
2U_j) \nonumber \\
&\equiv& n^o_{\tau}(E) P_{\tau}(E).
\end{eqnarray}
which defines the reduction factor $P_{\tau}(E)$ in the DOS; here we define
$U_j$ as the interaction between the given $\tau$ spin and the $j$-th S spin, and $E_{S_j}$ as
the unperturbed energy of this $S$ spin.
We would like to show that there is no appreciable reduction of the $\tau$ DOS by this
interaction, at least outside a window which is exponentially small in $g$. We proceed
by assuming a series of simplifying conditions, all leading to a reduction
of $P_{\tau}(E)$ below its actual value, and obtain our final result in form
of an inequality.

We first overestimate $n_S(E)=1/J_o$ for all $E<J_o$, neglecting the reduced
DOS at low energies. We then enumerate the $S$ spins according to their distance from
the $\tau$
impurity. Since there are $j$ impurities within a volume $(r_j)^3$,
the maximum interaction of the given $\tau$ TLS with the j'th S-TLS
is given by $U_j^{max}=g J_o/j$. We assume that all
interactions have this maximum value, taking the short distance cutoff to be
$\tilde{a} = 0$ and taking $c^{S \tau}_{ij}=1$. Under these assumptions,
for a given $E$ and a given $j$-th S-TLS we have $\Theta(E + E_{S_j} - 2U_j) =1$
if and only if $E_{S_j} > 2gJ_o/j - E$, with a probability $1 -  (2gJ_o/j - E)/J_o$
for $E < 2gJ_o/j$ (i.e. $j < 2gJ_o/E$) and unity otherwise.
Thus, for $E \ll g J_o$ we obtain

\begin{equation}
P_{\tau}(E) = \prod_j^{2 g J_o/E} \left(1-\frac{2 g J_o/j - E}{J_o} \right) .
\end{equation}
Defining $\epsilon \equiv E/2 g J_o$ we obtain

\begin{eqnarray}
P_{\tau}(\epsilon) = \prod_j^{1/\epsilon} \left[1-2 g \left(\frac{1}{j} -
\epsilon \right) \right] \;> \\ >\; \prod_j^{1/\epsilon} \left(1 - \frac{2 g}{j}\right) \; > \;
\prod_j^{1/\epsilon} 1/\left(1 + \frac{4 g}{j} \right) . \nonumber
\end{eqnarray}
Multiplying the denominators, and expanding in a series in $g$, we see that

\begin{equation}
\prod_j^{1/\epsilon} \left(1 + \frac{4 g}{j}\right) \;<\; 1 + \sum_k (4 g
\ln{1/\epsilon})^k.
\end{equation}
and therefore

\begin{equation}
P_{\tau}(\epsilon) \;>\;  1 - 4 g \ln{1/\epsilon}.
\label{Plower}
\end{equation}
i.e. $P_{\tau}(\epsilon) \approx 1$ for $\epsilon > \exp{(-1/8 g)}$.

Below we will argue that the S impurities are strongly gapped themselves,
so that the above small correction at $g^2 J_o < E_{\tau} < g J_o$ is
probably an overestimate.

Now consider the effect on $n_{\tau}(E)$ of the much weaker $\tau\tau$ interactions. By repeating
the same arguments as above, with $J^{SS}_o \equiv J_o$ replaced by $J^{S \tau}_o \equiv g
J_o$ and $J^{S \tau}_o$ replaced by $J^{\tau \tau}_o \equiv g^2 J_o$. Then Eq.(\ref{Plower})
is thus reproduced as an inequality. However, since the
$\tau$ impurities are not strongly gapped, one can follow through the same line of arguments with
$\approx$ instead of $>$, and conclude that there is a further reduction in the $\tau$-spin DOS by a
factor
\begin{equation}
\delta P_{\tau}(\epsilon') \;\approx\;  1 - c \ln{1/\epsilon'}.
\label{Pestimate}
\end{equation}
where $\epsilon' \equiv E/g^2 J_o$ and $c \approx g$.
Thus, the correction to the DOS appears at $E < g^2 J_o$, and is reduced by the small parameter $g$,
which is the ratio between the $\tau-\tau$ interaction and their energy disorder, and
is also the relevant small parameter in our theory. Experimentally, the dipole gap is
indeed seen at energy scales comparable to the $\tau-\tau$ interactions\cite{SRTO94},
and its magnitude is indeed considerably reduced.

\subsection{DOS of S impurities}

The calculation of the DOS of the S impurities is more subtle, since there is no small
parameter, and the result is of order unity. Furthermore,
the correlation between the interaction and $\tau$ energies is crucial, and must be
taken into account. We therefore solve the problem using a numerical simulation.
We begin by neglecting the $J^{\tau\tau}$ in Eq. (\ref{H-St}), since the values of $E_{\tau}$ are dictated by the $J^{S\tau}$ interaction. We are interested in the reduction of the S-TLSs resulting from the correlations with the $\tau$-TLSs which we denote by $P_S(E)$ and define by
\begin{eqnarray}
n_S(E) &=& n_S(\bar{E}) \prod_j \Theta(E + E_{\tau_j} -
2U_j) \nonumber \\
&\equiv& n_S(\bar{E}) P_S(E).
\end{eqnarray}
Here $\bar{E}$ is an energy a few times larger than $T_U$ and $U_j$ is the interaction
between a given $S$-TLS and $\tau_j$. We therefore use the following algorithm:

We consider a three dimensional cubic lattice with a given size and concentration, and
randomly distribute the impurities in the lattice. Each impurity has an S spin and a
$\tau$ spin. For a given $\tilde{a}$, the interaction is given by $U_{ij} = c_{ij}
S_i^z \tau_j^z J_0^{S \tau}/(R_{ij}^3 + {\tilde{a}}^3)$. We define $c'_{ij} \equiv
c_{ij} S_i^z$, where $c_{ij}$ is chosen randomly from a Gaussian distribution of width
unity and zero mean. We thus obtain $U_{ij}$ and $E_{\tau_j} \equiv \sum_i U_{ij}$ with
their essential dependence, but with no reference to the spin configuration of the $S$
TLSs. We then flip the $\tau$ spins where $E_{\tau} < 0$, to have a positive excitation
energy. The $U_{ij}$'s are accordingly redefined.
For each $S_i$ we then calculate $E^{min}_{S_i}$, the minimal $E$ satisfying
$\prod_j \Theta(E + E_{\tau_j} - 2U_{ij}) > 0$. We then obtain $P_S(E) = N(E^{min}_{S_i}
< E)/N(S)$, the ratio between the number of spins with $E^{min}_{S_i} < E$ to the total
number of spins. The results for x$=0.2$ are given in Fig. 2 in the main text, and for
x$=0.5$ in Fig. \ref{dos05}.

\subsection{random fields}
\label{SEC:random}

In the derivation above we neglected the explicit random fields $h^{\tau}, h^S$. This
requires special attention when calculating the DOS, $n_{\tau}(E), n_S(E)$. In
principle, the random fields add to the energies while keeping the interactions
unchanged, thus making the two particle stability condition easier to fulfil, reducing
the dipolar gap. Specifically, for $n_{\tau}(E)$ our argument above follows through
directly, and therefore the inequality in Eq. (\ref{Plower}) stays unchanged. For
$n_S(E)$ we have carried out the numerical calculations as described above in the
presence of random fields $h^{\tau}/\langle E_{\tau} \rangle \approx 0.3$ and
$h^{\tau}/\langle E_{\tau} \rangle \approx 1$ (where $\langle E_{\tau} \rangle$ is the
typical energy of a $\tau$ TLS in the absence of a random field). The dipolar gap of
the $S$ TLSs indeed becomes smaller, for $h^{\tau}/\langle E_{\tau} \rangle \approx
1$ the results are changed only slightly, whereas for $h^{\tau}/\langle E_{\tau}
\rangle \approx 0.3$ the quantitative change is appreciable, but the qualitative
behavior is unchanged, as can be seen in Fig. 2 of the main text.


\begin{thebibliography}{10}
\bibitem{tunnelingnote}
Tunneling is neglected in the calculatoin of the energy biases for symmteric and asymmetric TLSs because it is much smaller than the typical bias induced by the strong disorder, see discussion in App. \ref{microM}.
\end{thebibliography}
\end{document}